\documentclass[preprint,11pt]{aastex}
\usepackage{epsfig,rotating}
\tightenlines

\font\boldsym=cmmib10
\def    \balpha {{\hbox{\boldsym\char'013}}}    
\def    \bbeta  {{\hbox{\boldsym\char'014}}}
\def    \bgamma {{\hbox{\boldsym\char'015}}}
\def    \bA     {{\bf A}}
\def    \bE     {{\bf E}}
\def    \bJ     {{\bf J}}
\def    \bP     {{\bf P}}
\def    \bQ     {{\bf Q}}

\def    \ba     {{\bf a}}
\def    \be     {{\bf e}}
\def    \bk     {{\bf k}}
\def    \bn     {{\bf n}}

\def    \bq     {{\bf q}}
\def    \bx     {{\bf x}}

\def    \beq    {\begin{equation}}
\def    \eeq    {\end{equation}}

\begin{document}

\title{
	Propagation of Electromagnetic Waves on \\
	a Rectangular Lattice of Polarizable Points}

\author{D. Gutkowicz-Krusin}
\affil{Electro-Optical Sciences, Inc., 1 Bridge St., Irvington, NY 10533}
\and
\author{B.T. Draine}
\affil{Princeton University Observatory, Princeton, NJ 08544-1001}
\email{draine@astro.princeton.edu}

\begin{abstract}
We discuss the propagation of electromagnetic waves on a rectangular
lattice of polarizable point dipoles.
For wavelengths long compared to the lattice spacing, we obtain
the dispersion relation in terms of the lattice spacing and
the dipole polarizabilities.
We also obtain the dipole polarizabilities required for the lattice
to have the same dispersion relation as a continuum medium of
given refractive index $m$; our result differs from previous work by
Draine \& Goodman (1993).
Our new prescription can be used to assign dipole polarizabilities when
the discrete dipole approximation is used to study scattering by
finite targets.  Results are shown for selected cases.
\end{abstract}

\section{Introduction}

The discrete dipole approximation (DDA) provides a flexible and general
method to calculate scattering and absorption of light by objects of
arbitrary geometry \citep{Drai88,DrFl94}.
The approximation consists of replacing the continuum target of interest
by an array of polarizable points, which acquire oscillating electric
dipole moments in response to the electric field due to the incident
wave plus all of the other dipoles.
Broadly speaking, there are two criteria determining the accuracy of the
approximation:
\begin{itemize}
\item the interdipole separation $d$ should be small compared to the
	wavelength of the radiation in the material ($|m|kd < 1$,
	where $m$ is the complex refractive index, and
	$k=\omega/ c$ is the wavenumber {\it in vacuo});
\item the interdipole separation $d$ should be small enough to
	resolve structural dimensions in the target.
\end{itemize}
With modern workstations, it is now
feasible to carry out DDA calculations on targets containing up to
$\sim10^6$ dipoles \citep{DrFl94}.
In addition to calculation of scattering and absorption cross sections,
the DDA has recently been applied to computation of forces and torques
on illuminated particles \citep{DrWe96,DrWe97}

If the dipoles are located on a cubic lattice, then in the
limit where the interdipole separation $kd\rightarrow0$, 
the familiar Clausius-Mossotti
relation (see, e.g, \citet{Ja62}) can be used to determine the choice of
dipole polarizabilities required so that the dipole array will approximate
a continuum target with dielectric constant $\epsilon$.
\citet{Drai88} showed how this estimate for the dipole polarizabilities
should be modified to include radiative reaction corrections, and
\citet{DrGo93} derived the $O[(kd)^2]$ corrections
required so that an infinite cubic lattice would have the same
dispersion relation as a continuum of given dielectric constant.

Nearly all DDA calculations to date have assumed the dipoles to be located
on a cubic lattice.
If, instead, a rectangular lattice is used, 
it will still be possible
to apply FFT techniques to the discrete dipole approximation, in essentially
the same way as has been done for a cubic lattice \citep{GoDF91}.
However, the ability to use different lattice constants in different
directions might be useful
in representing certain target geometries.

The objective of the present report is to obtain the dispersion relation for
propagation of electromagnetic waves on a rectangular lattice of polarizable
points.
This dispersion relation will be expanded in powers of $(kd)$, where
$d$ is the characteristic interdipole separation.
For a lattice of specified dipole polarizabilities, this will allow us to
determine the dispersion relation for electromagnetic waves propagating on
the lattice when $(kd)^2\ll 1$.
Alternatively, if we require the lattice to propagate waves with a particular
dispersion relation, inversion of the lattice dispersion relation will
provide a prescription
for assigning dipole polarizabilities
when seeking to approximate a continuum material with a rectangular lattice
of polarizable points.

\section{Mode Equation for a Rectangular Lattice}

The problem of wave propagation on an infinite polarizable 
cubic lattice of point 
dipoles  has been solved previously by \citet[herafter DG93]{DrGo93}.
Here we describe the 
generalization of this problem to rectangular lattices. 

Consider an infinite rectangular lattice with lattice sites at 
\beq
\bx_\bn = (n_1 d_1,n_2 d_2,n_3 d_3)\, ,
\eeq
where the $n_i$ are integers and the 
lattice constants $d_i$ are, in general, all different. 
The density $n$ of dipoles is just
\beq
n = {1\over d_1d_2d_3} ~~~,
\eeq
and it will be convenient to define the characteristic lattice length scale
\beq
d \equiv (d_1d_2d_3)^{1/3} ~~~.
\eeq
Since the 
lattice is anisotropic, the polarizability $\balpha$ of the dipoles located 
at lattice sites is a tensor, i.e., the polarization is  
\beq
\bP(\bx)=\balpha \cdot \bE(\bx)\, .
\eeq
Thus the polarization vector is not, in general, parallel to the 
electric field or to the vector potential. 
From charge conservation we have 
\beq
\nabla\cdot\bJ = 0\, ,
\eeq
or, since 
\beq
\bJ = {\partial \bP \over \partial t} \, ,  
\label{eq:4}
\eeq
we obtain the transversality condition 
\beq
\nabla\cdot\bP = 0\, .
\label{eq:5}
\eeq
We assume that the dipole 
moment at the lattice site $\bx_\bn$ is 
\beq
\bP_\bn(t)= \bP_0(0) e^{i\bk\cdot\bx_\bn -i\omega t}\, .
\label{eq:6}
\eeq
With the Lorentz gauge condition
\beq
\nabla\cdot\bA + {1\over c}{\partial\Phi\over\partial t} = 0
\eeq
and equation (\ref{eq:4}), the vector potential
$\bA(\bx,t)$ satisfies the wave equation
\begin{eqnarray}
\nabla^2\bA + {\omega^2 \over c^2}\bA
	&=& -{4\pi\over c}\bJ
		\\
	&=& {4\pi i\omega\over c} \sum_{\bn} \bP_{\bn} \delta^3(\bx-\bx_\bn)
		\\
	&=&{4\pi i \omega \over c} \bP_0 e^{i\bk\cdot\bx}\sum_{\bn}
	\delta^3(\bx - \bx_\bn)~~~~~.\label{eq:10}
\end{eqnarray}

Consider now the unit cell centered on $\bx=0$.
Let $\bA_{other}$ and $\Phi_{other}$ be the potentials in this region due to
all dipoles {\it except} the dipole at $\bx=0$.
Thus
\beq
\bE_{other}(\bx,t) =-{1\over c}{\partial\bA_{other}\over\partial t}-
\nabla\Phi_{other}
= {i\omega\over c}\bA_1(\bx,t)
\eeq
where
\beq
\bA_1(\bx,t) \equiv \bA_{other} + {c^2\over\omega^2}\nabla(\nabla\cdot\bA_{other})
\label{eq:12}~~~~.
\eeq
Thus
\beq
\bP_0(0) = { i \omega \over c}\, \balpha \cdot \bA_1(\bx=0,t=0)
\label{eq:13}~~~~.
\eeq
The vector potential can be written
\beq
\bA(\bx,t)= e^{i\bk\cdot\bx-i\omega t}\sum_{\bn}\ba_{\bn}e^{i \bq\cdot\bx}
\label{eq:14}
\eeq
where 
\beq
\bq \equiv \bq(\bn) = 2 \pi \, \bigg ( {n_1 \over d_1}, {n_2 \over d_2}, {n_3 \over d_3} \bigg )
\eeq
define a lattice which is reciprocal (e.g., \cite{AM76}) 
to the original rectangular lattice on
which the dipoles are located.
Equations (\ref{eq:10}) and (\ref{eq:14}), and the identity
\beq
\sum_{\bn}\delta^3(\bx-\bx_\bn)= {1 \over d_1 d_2 d_3} \sum_{\bn}e^{i\bq\cdot\bx}
\label{eq:16}
\eeq
yield
\beq
\ba_{\bn} = {4\pi\omega^2\over c^2}\, {1\over d_1 d_2 d_3 }\, \balpha \cdot\bA_1(0,0)
\left[|\bk+\bq|^2-{\omega^2\over c^2}\right]^{-1}
~~~~.
\eeq
The field $\bA_{self}(\bx,t)$ due to the dipole
at $\bx=0$ is
\beq
\bA_{self}(\bx,t) = {-i\omega\over c}\bP_0 {e^{i\omega r/c} \over r}
= {\omega^2 \over c^2}\, \balpha \cdot \bA_1(0,0)
{e^{i\omega r/c} \over r} e^{-i\omega t}
~~~~.
\eeq
With the identity
\beq
{e^{i\omega r/c}\over r} = {e^{i\bk\cdot\bx} \over 2\pi^2}
\int d^3k^\prime {e^{i\bk^\prime\cdot\bx}\over
|\bk^\prime+\bk|^2-\omega^2/c^2}
\eeq
we obtain
\begin{eqnarray}
\bA_{other} = {4\pi\omega^2\over c^2}
{1\over d_1 d_2 d_3 }\, 
e^{i\bk\cdot\bx-i\omega t}\,
\balpha \cdot \bA_1(0,0)
\bigg[
\sum_\bn{e^{i\bq\cdot\bx} \over |\bk+\bq|^2-\omega^2/c^2}
~-\nonumber
\\
{ d_1 d_2 d_3 \over 8\pi^3} \int {d^3k^\prime}
{e^{i\bk^\prime\cdot\bx} \over |\bk+\bk^\prime|^2-\omega^2/c^2}
\bigg]
\label{eq:20}~~~.
\end{eqnarray}
We seek $\bA_1$. 
Substituting (\ref{eq:20}) into (\ref{eq:12}), 
we obtain after some algebra:
\begin{eqnarray}
A_{1,i}(\bx,t) =  {4 \pi \over d_1 d_2 d_3} \sum_{j=1}^3 \sum_{l=1}^3 
\biggl[
	\sum_\bn
	\left(
		{\omega^2\over c^2}\delta_{il}-(k_i+q_i)(k_l+q_l)
	\right)
	{e^{i(\bk+\bq)\cdot\bx}\over|\bk+\bq|^2-\omega^2/c^2}
	~-
\nonumber
\\
	{d_1 d_2 d_3 \over 8 \pi^3} \int{d^3k^\prime }
	\left(
		{\omega^2\over c^2}\delta_{il}-(k_i+k_i^\prime)
		(k_l+k_l^\prime)
	\right)
	{e^{i(\bk+\bk^\prime)\cdot\bx}\over |\bk+\bk^\prime|^2-\omega^2/c^2}
\biggr]
e^{-i\omega t}\,
\alpha_{lj} A_{1,j}(0,0) \, . 
\label{eq:21}
\end{eqnarray}
For $\bx = 0$ and $t = 0$ and given values of $\alpha_{ij}$, 
equation (\ref{eq:21}) 
is the mode equation, i.e., it  
allows determination of the dispersion relation $\bk = \bk (\omega )$. 
However, in the present case, we know the dispersion relation, and wish 
to determine values of $\alpha_{ij}$ such that the known dispersion 
relation and equation (\ref{eq:21}) are satisfied simultaneously. Another 
relation that must be satisfied follows from (\ref{eq:5}), (\ref{eq:6}), 
(\ref{eq:13}), and (\ref{eq:16}):
\beq
\bk \cdot \bP_0(0) = \bk \cdot \balpha \cdot \bA_1(0,0) = 0\, .  
\eeq
Hence, if $\be$ is a unit vector in the direction of $\bA_1$, then 
\beq
\sum_{j=1}^3 \sum_{l=1}^3 k_l \, \alpha_{lj} \, e_j = 0\, .  
\eeq
At this point it is convenient to define the following dimensionless 
quantities:
\beq
\tilde q \equiv  {2 \pi \over d}
= {2 \pi \over (d_1 d_2 d_3)^{1/3} }
\label{eq:24}
\eeq
\beq
\nu \equiv {\omega \over \tilde q c}\, ;
\label{eq:25}
\eeq
\beq
\bbeta \equiv {\bk \over \tilde q}\, ;
\label{eq:26}
\eeq
\beq
\bgamma \equiv {4 \pi \over d_1 d_2 d_3 } \, \balpha \, ;
\label{eq:27}
\eeq
\beq
\bQ \equiv {\bq \over \tilde q} = 
	(n_1{d\over d_1},  
	n_2{d\over d_2},  
	n_3{d\over d_3})\, , 
\label{eq:28}
\eeq
The vectors $\bQ$ define a reciprocal lattice
(e.g., \cite{AM76}).

Using these dimensionless quantities, the mode equation (\ref{eq:21}) may 
be written
\beq
\sum_{j=1}^3 \sum_{l=1}^3 M_{il}\, \gamma_{lj}\, e_j = e_i\, ,
\label{eq:mode_equation}
\eeq
where
\beq
M_{ij} \equiv
	\sum_\bn
	{\nu^2\delta_{ij}-(\beta_i+Q_i)(\beta_j+Q_j)
	\over
	|\bbeta+\bQ|^2-\nu^2}
	-
	\int d^3\bbeta^\prime
	{\nu^2\delta_{ij}-(\beta_i+\beta_i^\prime)(\beta_j+\beta_j^\prime)
	\over
	|\bbeta+\bbeta^\prime|^2-\nu^2}
~~~,
\label{eq:31}
\eeq
and equation (25) becomes
\beq
\sum_{i=1}^3 \sum_{j=1}^3 \beta_i \, \gamma_{ij} \, e_j = 0\, .  
\label{eq:32}
\eeq
It should be noted that the mode equation for a cubic 
lattice has the same form as equation (\ref{eq:mode_equation}).
The matrix elements $M_{ij}$ in equation (\ref{eq:31}) 
also have the same form as the matrix elements for a cubic lattice 
but with the vector $\bn$ 
replaced by $\bQ$ for a rectangular lattice. 
Since the matrix elements for a cubic lattice have been calculated previously, 
it is convenient to rewrite (\ref{eq:31}) as 
\beq
M_{ij} = M_{ij}^C + T_{ij} \, ,
\label{eq:33}
\eeq
where $M_{ij}^C$ is the matrix element for a cubic lattice, and $T_{ij}$ is 
\beq
T_{ij} \equiv
	\sum_\bn
	{\nu^2\delta_{ij}-(\beta_i+Q_i)(\beta_j+Q_j)
	\over
	|\bbeta+\bQ|^2-\nu^2}
	-
	\sum_\bn
	{\nu^2\delta_{ij}-(\beta_i+n_i)(\beta_j+n_j)
	\over
	|\bbeta+\bn|^2-\nu^2}
~~~.
\label{eq:34}
\eeq

\section{Dispersion Relation in the Long-Wavelength Limit}

For a given polarizability tensor $\bgamma$, the mode equation (\ref{eq:mode_equation})
allows one to determine the dispersion relation $\bbeta = \bbeta (\nu)$.
In the present case, the dispersion relation is known:
\beq
\bbeta = m\nu \ba \, ,
\eeq
where $m=\epsilon^{1/2}$
is the (complex) refractive index and $\ba$ is a unit vector 
in the direction of propagation, which is fixed. 
From equation (\ref{eq:32}) for 
a rectangular lattice, the polarization rather than the vector potential 
is perpendicular to the direction of propagation, i.e., 
\beq
\sum_{i=1}^3 \sum_{j=1}^3 a_i \, \gamma_{ij} \, e_j = 0\, .  
\label{eq:36}
\eeq
Therefore, the mode equation (\ref{eq:mode_equation}) can be used to determine dipole 
polarizabilities $\gamma_{ij}$ such that 
equation (\ref{eq:36}) is satisfied.  

We seek the dipole polarizabilities in the limit 
$\nu^2 \ll 1$. In this limit, one can write
\beq
\gamma_{ij} = \gamma_{ij}^{(0)} + \gamma_{ij}^{(1)} + O(\nu^4)
~~~,
\eeq
\beq
M_{ij}^C=M_{ij}^{C(0)} + M_{ij}^{C(1)} + O(\nu^4)
~~~,
\eeq
\beq
T_{ij}=T_{ij}^{(0)} + T_{ij}^{(1)} + O(\nu^4)
~~~,
\eeq
\beq
\be = \be^{(0)} + \be^{(1)} + O(\nu^4)
~~~,
\eeq
where superscripts 0 and 1 designate the zero frequency limit and
the leading order corrections, respectively. 

The low-frequency expansion of the matrix element $M_{ij}$ for 
a cubic lattice was given by DG93 and is:
\beq
M_{ij}^{C(0)} = {m^2+2\over3(m^2-1)}\delta_{ij} - {m^2\over m^2-1}a_i a_j
~~~,
\label{eq:41}
\eeq

\beq
M_{ij}^{C(1)} = \nu^2\delta_{ij}
\left[c_1+m^2c_2(1-3a_i^2)\right] + m^2\nu^2(1-\delta_{ij})c_3a_i a_j-
{4\pi^2i\over3}\delta_{ij}\nu^3
~~~,
\label{eq:42}
\eeq
where
\begin{eqnarray}
c_1 &=& -5.9424219... ~~~,\nonumber  \\
c_2 &=& 0.5178819... ~~~,\nonumber  \\
c_3 &=& 4.0069747... ~~~. \nonumber
\end{eqnarray}
From equation (\ref{eq:34}), one finds 
\beq
T_{ij}^{(0)} = \delta_{ij} R_0 (i)
~~~, 
\label{eq:43}
\eeq
where
\beq
R_0(i)=\sum_\bn
	 { n_i^2 \over |\bn|^2 } - 
	\sum_\bn
	 { Q_i^2 \over |\bQ|^2 }
\label{eq:R_0}~~~.
\eeq
Note that $\sum_{i=1}^3R_0(i)=0$, so that there are only two independent
sums which require evaluation.
Numerical evaluation is discussed in the Appendix. 
\begin{eqnarray}
T_{ij}^{(1)} &=& m^2 \nu^2 a_ia_j \left[ R_1 - 2 R_2(i) - 2 R_2(j)
               + 8 R_3(i,j) \right]  \nonumber\\
             &-& \nu^2 \delta_{ij} 
  \left[ R_1 + (m^2 -1) R_2(i) + 8m^2 a_i^2 R_3 (i,i) 
  - 4m^2 \sum_{l=1}^3 a_l^2 R_3(i,l) \right] \, , 
\label{eq:44}
\end{eqnarray}
where
\beq
R_1 =  	\sum_\bn { 1 \over |\bn|^2 } - 
	\sum_\bn { 1 \over |\bQ|^2 }  \, ,
\eeq
\beq
R_2(i) =  \sum_\bn { n_i^2 \over |\bn|^4 } - 
	\sum_\bn { Q_i^2 \over |\bQ|^4 }  \, ,
\eeq
\beq
R_3(i,j) =  \sum_\bn { n_i^2 n_j^2  \over |\bn|^6 } - 
	\sum_\bn { Q_i^2 Q_j^2  \over |\bQ|^6 }  \, .
\label{eq:R_3}
\eeq
In the limit of cubic lattice, $Q_i \rightarrow n_i$, and all the 
sums $R_n$ in equations (\ref{eq:R_0}) - (\ref{eq:R_3}) vanish. 
For a general rectangular 
lattice, these sums must be evaluated for each set of new 
lattice constants. Care must be used in evaluation of these 
sums as described in the next section. 
We note that $R_1$ and $R_2$ may be obtained from $R_3$:
\beq
R_1 = \sum_{i=1}^3\sum_{j=1}^3 R_3(i,j)
\label{eq:R_1fromR_3}~~~;
\eeq
\beq
R_2(i) = \sum_{j=1}^3 R_3(i,j)
\label{eq:R_2fromR_3}~~~.
\eeq
Since $R_3(i,j)=R_3(j,i)$, six independent quantities
suffice to determine $R_1$, $R_2$, and $R_3$.

\subsection{Static Limit}

In the static limit the polarizability tensor 
$\gamma_{ij}^{(0)}$ is diagonal, 
and the mode equation and transversality 
condition become:
\beq
\sum_{j=1}^3 \left [ M_{ij}^{C(0)} + T_{ij}^{(0)} \right ]\, 
\gamma_{jj}^{(0)}\, e_j^{(0)} = e_i^{(0)}\, ,
\label{eq:48}
\eeq
\beq
\sum_{i=1}^3 a_i \, \gamma_{ii}^{(0)} \, e_i^{(0)} = 0\, .  
\label{eq:49}
\eeq
Substituting (\ref{eq:41}) and (\ref{eq:43}) into (\ref{eq:48}) 
and taking into account condition (\ref{eq:49}) 
one gets 
\beq
\left[ {m^2 + 2 \over 3(m^2 -1) } + R_0(i) \right]\, 
\gamma_{ii}^{(0)}\, e_i^{(0)} = e_i^{(0)}\, .
\label{eq:50}
\eeq
For $e_i^{(0)} \neq 0$, the static dipole polarizability for a rectangular 
lattice is 
\beq
\gamma_{ii}^{(0)} = { \gamma_{CM} \over 1 + \gamma_{CM} R_0(i)} \, ,
\label{eq:51}
\eeq
where the Clausius-Mossotti polarizability is
\beq
\gamma_{CM}  \equiv  { 3(m^2 -1 ) \over m^2 + 2 } \, .
\eeq
It is evident from equation (\ref{eq:51}) that the static 
polarizability for a cubic lattice 
reduces to the Clausius-Mossotti result. 


If $e_i^{(0)} = 0$ in one of the directions, then the polarizability 
in this direction 
is undetermined by equation (\ref{eq:50}). 
However, since the result (\ref{eq:51}) is valid 
for values of $e_i^{(0)}$ arbitrarily close to zero, we take (\ref{eq:51}) 
to be the 
solution for all directions of the electric field subject to 
equation (\ref{eq:49}).  

\subsection{Finite Wavelength Corrections}

We now proceed to obtain the leading corrections for small $\nu$.
The first order mode equation and transversality 
condition are:
\beq
 \sum_{j=1}^3 \sum_{l=1}^3 M_{il}^{(1)} \gamma_{lj}^{(0)}\, e_j^{(0)} 
 + \sum_{j=1}^3 \sum_{l=1}^3 M_{il}^{(0)} \left[ \gamma_{lj}^{(1)}\, e_j^{(0)} 
 +  \gamma_{lj}^{(0)}\, e_j^{(1)} \right]  
= e_i^{(1)}\, , 
\label{eq:53}
\eeq
\beq
\sum_{i=1}^3 \sum_{j=1}^3 a_i \, \left[ \gamma_{ij}^{(1)} \, e_j^{(0)} 
+  \gamma_{ij}^{(0)}\, e_j^{(1)} \right ]  = 0\, .  
\label{eq:transversality}
\eeq
From equations (\ref{eq:41}), (\ref{eq:43}), (\ref{eq:50}), 
and (\ref{eq:transversality}) one gets 
\begin{eqnarray}
 \sum_{j=1}^3 \sum_{l=1}^3 M_{il}^{(0)} \left[ \gamma_{lj}^{(1)}\, e_j^{(0)} 
 +  \gamma_{lj}^{(0)}\, e_j^{(1)} \right]  
&=& \sum_{j=1}^3 \left[ {m^2 + 2 \over 3(m^2 -1) } + R_0(i) \right]\, 
 \left[ \gamma_{ij}^{(1)}\, e_j^{(0)} 
 +  \gamma_{ij}^{(0)}\, e_j^{(1)} \right] \nonumber \\
&=& \sum_{j=1}^3 {\gamma_{ij}^{(1)} \over \gamma_{ii}^{(0)} } e_j^{(0)} + e_i^{(1)} \, . 
\end{eqnarray}
Therefore, equation (\ref{eq:53}) becomes 
\beq
\label{eq:60}
\sum_{j=1}^3 
\left[ 
	{\gamma_{ij}^{(1)} \over \gamma_{ii}^{(0)}} 
	+ 
	M_{ij}^{(1)} \gamma_{jj}^{(0)}
\right]
\,e_j^{(0)}=0 ~~~ .  
\label{eq:the_equation}
\eeq
It is convenient to write $M_{ij}^{(1)}$ in the form 
\beq
M_{ij}^{(1)} = \nu^2 \delta_{ij} L_{ii} + 
m^2\nu^2 a_i a_j 
K_{ij}						
~~~,
\eeq
where, from (\ref{eq:42}) and (\ref{eq:44})
\begin{eqnarray}
L_{ii} &=&
c_1+m^2c_2(1-3a_i^2) - m^2 c_3a_i^2 -
{4\pi^2i\over3}\nu \nonumber\\ 
&-& R_1 -  (m^2 -1) R_2(i) - 8m^2 a_i^2 R_3 (i,i) 
  + 4m^2 \sum_{l=1}^3 a_l^2 R_3(i,l)
\end{eqnarray}
and 
\beq
K_{ij} = 
c_3 + R_1 -2 R_2(i) -2 R_2(j) + 8 R_3(i,j)
\, .
\eeq
We now observe that one can have the square bracket in equation 
(\ref{eq:the_equation})
vanish for each $j$ if the first-order correction to the polarizability is
\begin{eqnarray}
\gamma_{ij}^{(1)}
&=&-\gamma_{ii}^{(0)}\gamma_{jj}^{(0)}M_{ij}^{(1)}\\
&=&
-\nu^2 \gamma_{ii}^{(0)} \gamma_{jj}^{(0)} \left[ L_{ii} \delta_{ij} + 
m^2  a_i a_j K_{ij} \right ] \, .
\label{eq:gamma_ij^(1)}
\end{eqnarray}
Equation (\ref{eq:gamma_ij^(1)}) 
makes it explicit that the frequency-dependent 
term in the dipole polarizability depends on the direction 
of propagation of the wave.

We observe, however, that this solution is not unique.
In fact, the transversality condition (\ref{eq:transversality})
permits the more general solution

\beq
\gamma_{ij}^{(1)} =
-\nu^2 \gamma_{ii}^{(0)} \gamma_{jj}^{(0)} \left[ L_{ii} \delta_{ij} + 
m^2  a_i a_j \left( K_{ij} + c_4 \right) \right ]
\eeq
where $c_4$ could, in principle, be a function of
$m^2$, the direction of propagation, or the polarization.
We note, however, that if $c_4\neq 0$ then 
eq. (\ref{eq:the_equation}) is
no longer satisfied term by term, but only after summation.

Since we appear to have freedom in the choice of $c_4$,
numerical experiments using
different choices for $c_4$ could be used to see what value of $c_4$
appears to be optimal for 
scattering calculations using the discrete dipole approximation..

\subsection{Special Case of a Cubic Latttice}

Propagation of electromagnetic waves on a cubic lattice has
been discussed by DG93.  In their analysis
it was assumed that if the dielectric material is isotropic,
the polarizability $\gamma_{ij}=\gamma_{ij}^{(0)}+\gamma_{ij}^{(1)}$
can be taken to be isotropic:
$\gamma^{(0)}_{ij}=\gamma^{(0)}\delta_{ij}$
and 
$\gamma^{(1)}_{ij}=\gamma^{(1)}\delta_{ij}$.
With this assumption, contraction of eq. (\ref{eq:60})
with $e_i^{(0)}$ leads to
\beq
\label{eq:alpha_DG93}
\gamma^{(1)}_{ij} = -\left[\gamma_{jj}^{(0)}\right]^2\nu^2
\delta_{ij}
\left[c_1+m^2c_2-m^2(3c_2+c_3)S - \frac{4\pi^2}{3}i\nu\right]
~~~,
\eeq
\beq
S\equiv\sum_{j=1}^3 \left(a_je_j^{(0)}\right)^2
~~~.
\eeq
as in eq.(4.14) of \cite{DrGo93}.
However, we have seen above that for a cubic lattice 
(for which $K_{ij}=c_3$), $\gamma_{ij}^{(1)}$ can be
made diagonal but not isotropic: if we choose $c_4=-c_3$ so
that $\gamma_{ij}^{(1)}$ is diagonal, we obtain
\beq
\label{eq:alpha_cubic}
\gamma^{(1)}_{ij} = -\left[\gamma_{jj}^{(0)}\right]^2\nu^2
\delta_{ij}
\left[c_1+m^2c_2-m^2(3c_2+c_3)a_j^2 - \frac{4\pi^2}{3}i\nu\right]
~~~.
\eeq
which differs from the result of \citet{DrGo93} since $a_j^2\neq S$,
except for the special case of propagation in the (1,1,1) direction,
for which $a_i^2=S=1/3$.

\begin{figure}[h]
\centerline{\epsfig{
    file=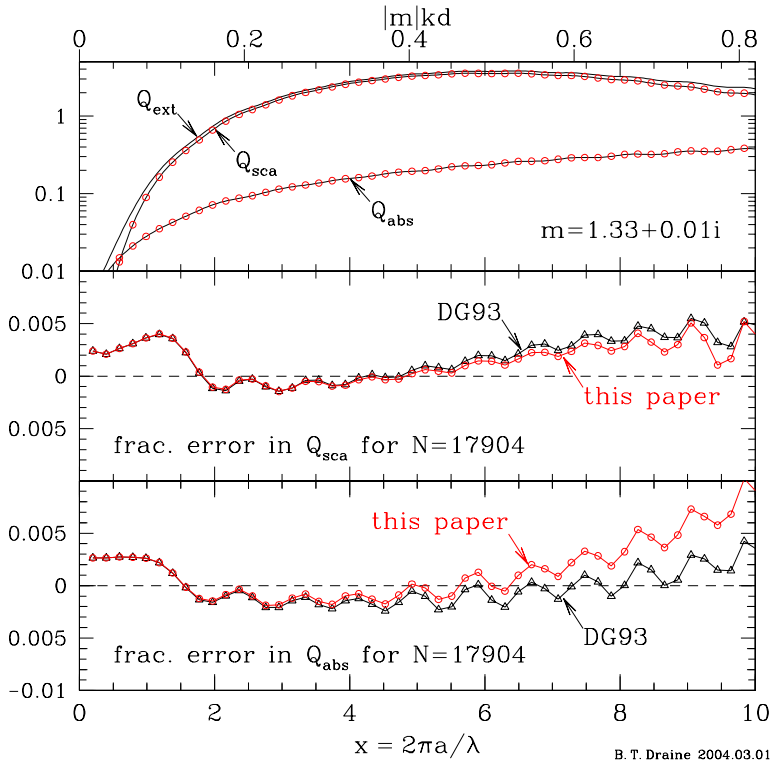,
    width=4.0in,
    angle=0}}
\caption{\label{fig:m=1.33+0.01i}\footnotesize
  DDA calculations of scattering and absorption by a sphere with
  refractive index $m=1.33+0.01i$.
  Upper panel shows $Q_{\rm sca}$, $Q_{\rm abs}$, and $Q_{\rm ext}\equiv
  Q_{\rm sca}+Q_{\rm abs}$ calculated using Mie theory (solid curves) and
  using the DDA with $N=17904$ and the present polarizabilities.
  Middle and lower panels show fractional errors for $Q_{\rm sca}$ and
  $Q_{\rm abs}$.
  Circles: using dipole polarizabilities from this paper 
  (eq.(\ref{eq:alpha_cubic}).
  Triangles: using dipole polarizabilities from Draine \& Goodman (1993);
  }
\end{figure}
\begin{figure}[h]
\centerline{\epsfig{
    file=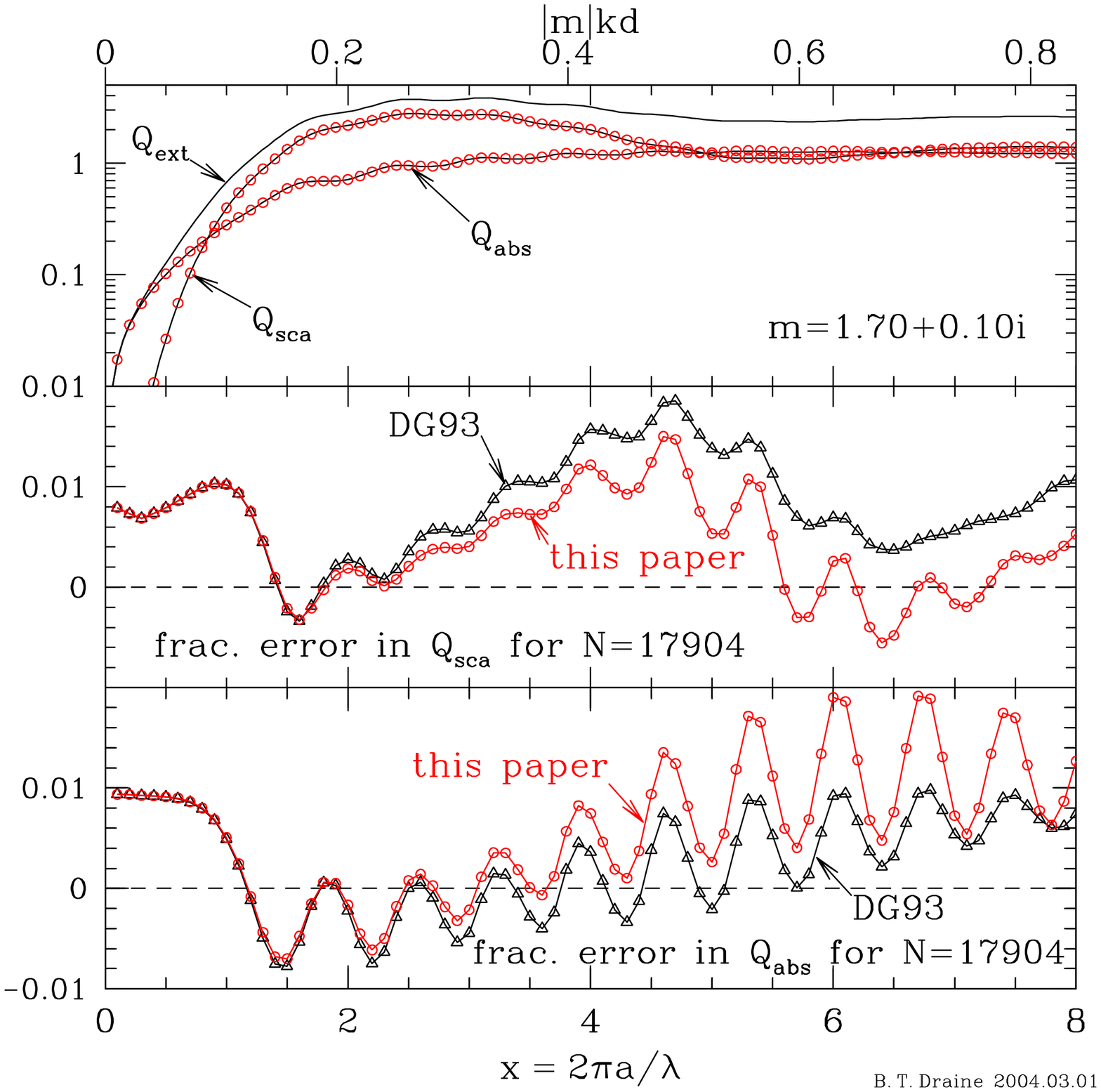,
    width=4.0in,
    angle=0}}
\caption{\label{fig:m=1.70+0.10i}\footnotesize
  As in Figure \ref{fig:m=1.33+0.01i}, but for $m=1.7+0.1i$.
  }
\end{figure}

\section{DDA Scattering Tests}

We now compare discrete dipole approximation (DDA) calculations of
scattering and absorption using (1) the polarizability prescription
(\ref{eq:alpha_DG93}) from DG93 and (2) the present result
(\ref{eq:alpha_cubic}).
We consider spheres where we can use Mie theory to
easily calculate the exact result for
comparison.
Let $C_{\rm sca}$ and $C_{\rm abs}$ be the cross sections for scattering
and absorption, and let $Q_{\rm sca}\equiv C_{\rm sca}/\pi a^2$ 
and
$Q_{\rm abs}\equiv C_{\rm abs}/\pi a^2$ 
be the dimensionless scattering
and absorption efficiency factors for a sphere of radius $a$.

Figure \ref{fig:m=1.33+0.01i} shows
$Q_{\rm sca}$ and
$Q_{\rm abs}$ 
calculated using the DDA with a target
consisting of $N=17904$ dipoles,
with   
radius $a=a_{\rm eff}\equiv (3N/4\pi)^{1/3}d$.
The DDA calculations used a cubic lattice, and 
were carried out using the public domain
code DDSCAT.6.1 \citep{DrFl04}.
Calculations were done for values of the scattering parameters 
$x\equiv 2\pi a/\lambda$
satisfying $|m|kd < 0.8$, as this appears to be a
good validity criterion for DDA calculations.
The upper panel shows scattering and absorption efficiencies
calculated with Mie theory and using the DDA with the present prescription
for polarizabilities $\balpha$.
The middle and lower panels show fractional errors 
$[Q({\rm DDA})/Q({\rm Mie}) -1]$
of the DDA
scattering and absorption cross sections.

Because 
the DDA target (dipoles on a cubic lattice)
is not rotationally symmetric,
the scattering problem was solved for 12 different
incident directions, in each case for two orthogonal polarizations.
The cross sections shown in Fig. \ref{fig:m=1.33+0.01i} are averages over
the 12 incident directions and two incident polarizations.

In the limit $x\rightarrow 0$,
identical results are obtained 
for both polarizability prescriptions; this is expected since
they differ only in terms of $O(|m|^2 (kd)^2)$.  Note that even
at $x=0$ (i.e., $kd=0$) -- for which the polarizabilities are exactly
given by the Clausius-Mossotti polarizabilities -- the DDA calculation
has a nonzero error even as $x\rightarrow0$ and $kd\rightarrow 0$.
This is because (1) the dipole array only approximates a spherical
geometry and (2) when the ``lattice dispersion relation'' polarizabilities
are used, the DDA does not accurately model the ``screening'' of
the target material close to the surface
by the polarization at the target surface.
For modest dielectric functions, and sufficient numbers of dipoles,
the errors are small -- for the case shown in 
Figure \ref{fig:m=1.33+0.01i}
the fractional errors in the $x\rightarrow0$ limit are $<0.3\%$.

As $x$ increases,
the two prescriptions result in differences in computed values of
$Q_{\rm sca}$ and $Q_{\rm abs}$, although the differences are
small, and comparable in magnitude to the errors in the
$x\rightarrow0$ limit.
In Figure \ref{fig:m=1.33+0.01i} our new prescription appears to
result in small improvements in accuracy (relative to DG93) for 
$Q_{\rm sca}$
for $|m|kd<0.8$, and likewise for $Q_{\rm abs}$ for $|m|kd<0.5$.
For $0.5<|m|kd<0.8$, however,
the DG93 prescription appears to give more accurate results for
$Q_{\rm abs}$.  Note, however, that for both polarizability prescriptions
the fractional errors in $Q_{\rm sca}$ and $Q_{\rm abs}$
for $|m|kd \approx 0.5$
are of the same order as the fractional errors for $kd\rightarrow 0$,
so it is not clear that the errors should be attributed to the
lattice dispersion relation -- the errors in $Q_{\rm sca}$ and
$Q_{\rm abs}$ may instead
be associated with errors in the computed polarization field near
the surfaces, arising from the fact that in the DDA 
the dipoles in the surface
layer are imperfectly ``screened'' from the external field by their
neighbors.

\begin{figure}[h]
\centerline{\epsfig{
    file=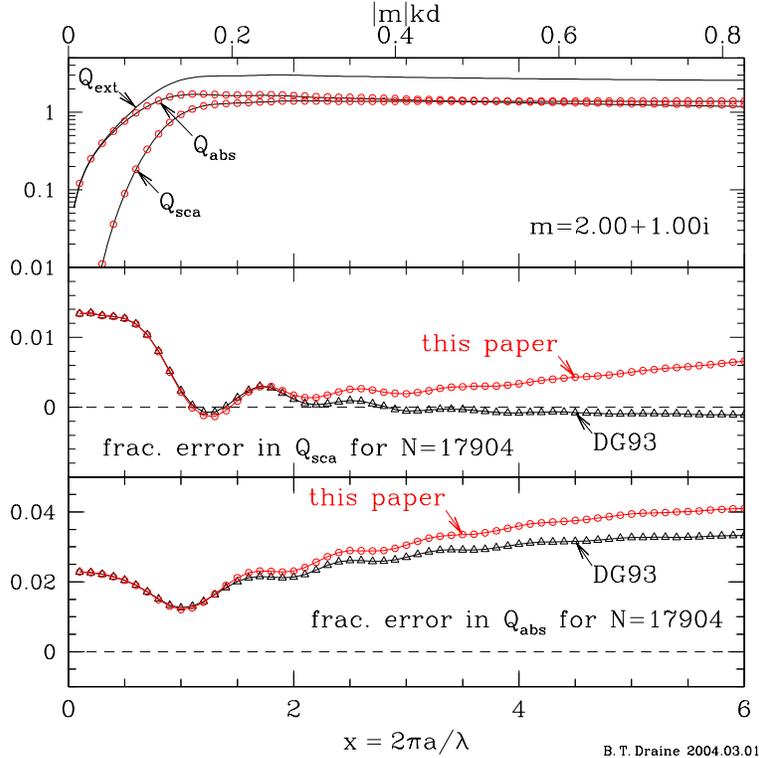,
    width=4.0in,
    angle=0}}
\caption{\label{fig:m=2.00+1.00i}\footnotesize
  As in Figure \ref{fig:m=1.33+0.01i}, but for $m=2+i$.
  }
\end{figure}
\begin{figure}[h]
\centerline{\epsfig{
    file=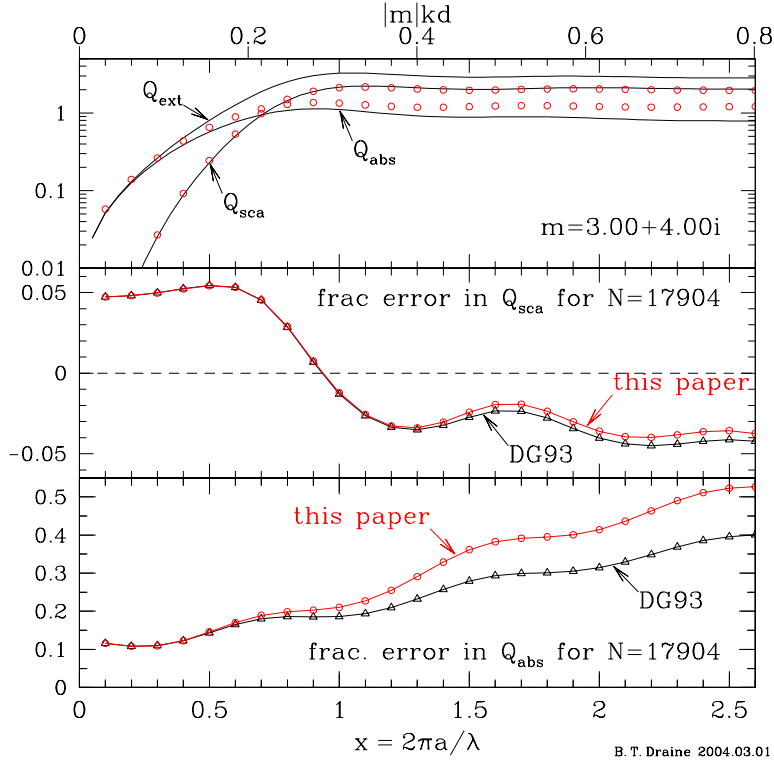,
    width=4.0in,
    angle=0}}
\caption{\label{fig:m=3.00+4.00i}\footnotesize
  As in Figure \ref{fig:m=1.33+0.01i}, but for $m=3+4i$.
  }
\end{figure}

\begin{figure}[h]
\centerline{\epsfig{
    file=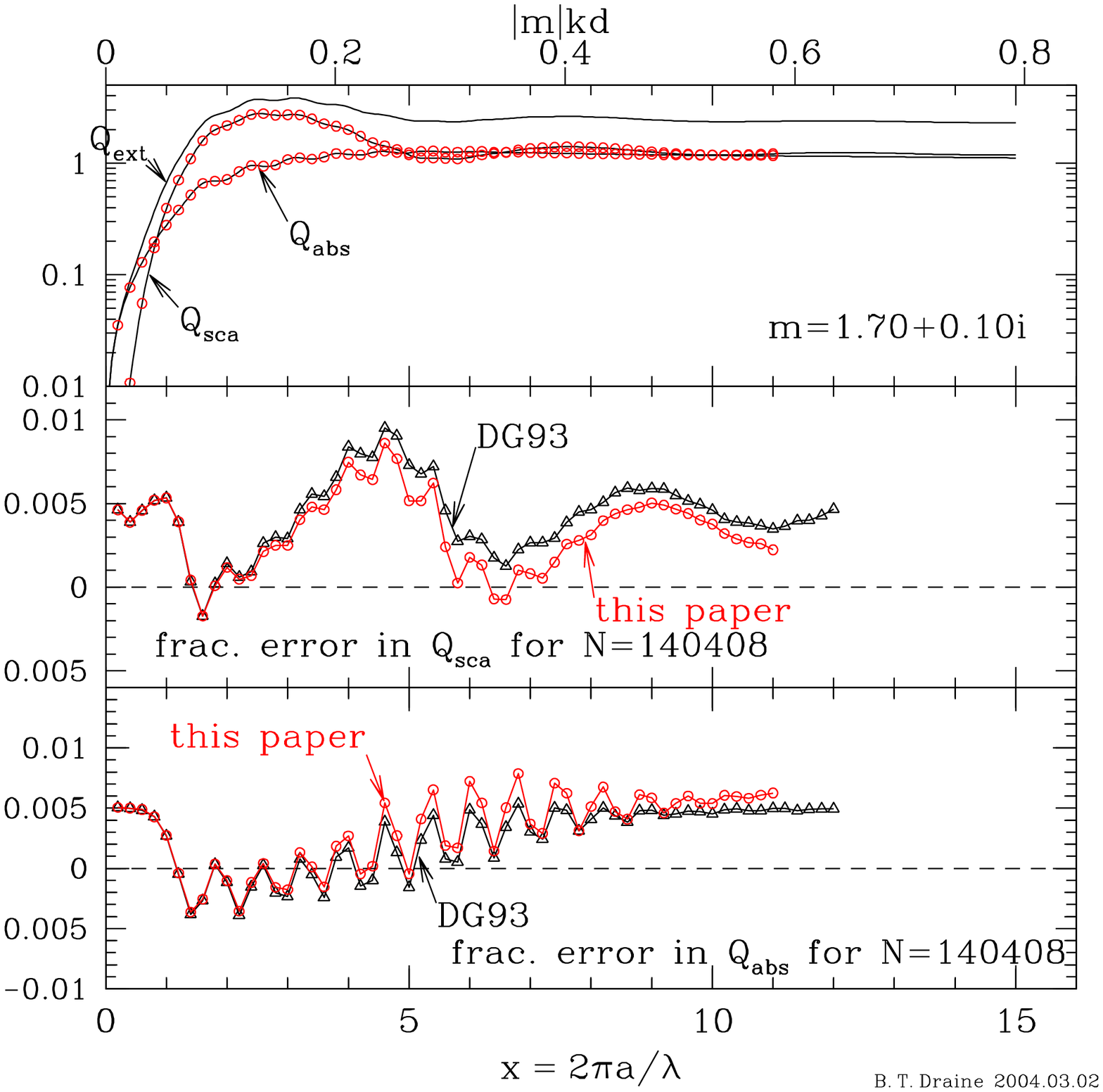,
    width=4.0in,
    angle=0}}
\caption{\label{fig:m=1.70_0.10ib}\footnotesize
  As in Figure \ref{fig:m=1.70+0.10i}, but for $N=140408$ dipoles.
  }
\end{figure}

For $|m|kd < 0.8$, the errors in $Q_{\rm sca}$
using the new polarizabilities are of order $\sim 1\%$ 
for  
$m=1.33+0.01i$ (Figure \ref{fig:m=1.33+0.01i}),
$m=1.7+0.1i$ (Figure \ref{fig:m=1.70+0.10i}) and
$m=2+i$ (Figure \ref{fig:m=2.00+1.00i}),
and  of order $\sim5\%$ for
$m=3+4i$ (Figure \ref{fig:m=3.00+4.00i}).
In all four cases, for wavelengths satisfying $|m|kd<0.8$,
the magnitude of the error in $Q_{\rm sca}$ 
is of the same order as the error
at zero frequency, suggesting that it is mainly due to the
surface.

The errors in $Q_{\rm abs}$ tend to be somewhat larger than
for $Q_{\rm sca}$, especially
for large values of ${\rm Im}(m)$.
For $m=3+4i$ (Figure \ref{fig:m=3.00+4.00i}), the new polarizabilities
result in a fractional error of $\sim50\%$ for $x=2.6$ and
$N=17904$ dipoles.

Figure \ref{fig:m=1.70_0.10ib} shows results calculated for 
$m=1.7+0.1i$ spheres represented by $N=140408$ dipoles -- 8 times as
many as in Fig. \ref{fig:m=1.70_0.10i}.  Note that the fractional
error for $x\rightarrow 0$ is now only 0.5\% -- half as large as
for $N=17904$.  This can be ascribed to the fact that the fraction of
the dipoles in the surface monolayer was reduced by a factor of 2.
Note also that the errors for finite $x$ are also about a factor of 2 smaller
than for $N=17904$.  These results appear to confirm the suspicion that
the errors arise primarily from the surface layer.

\citet{RCB02} recently proposed a method for
assigning dipole polarizabilities that takes into account the effects
of target geometry, resulting in dipole polarizabilties that are
location-dependent even when the target being approximated
has homogeneous composition.
\citet{CoDr04} have extended this treatment by adding 
finite-wavelength corrections to the
 ``surface-corrected'' polarizability prescription, and used
DDA calculations for spheres and ellipsoids to show that accurate results
for $Q_{\rm sca}$ and $Q_{\rm abs}$ 
can be obtained for modest numbers of dipoles even when the
target material has a large refractive index $|m|\geq 5$.
The current calculations does {\it not} include such ``surface corrections'',
-- the prescription (\ref{eq:alpha_cubic}) depends only on
local properties -- 
so it comes as no surprise that there are systematic errors even
in the limit $x\rightarrow0$.

\citet{CoDr04} show that the error can be greatly reduced if the
polarizabililites are assigned in a way that allows for both 
target geometry (i.e., surface corrections) and the finite
wavelength corrections obtained from the lattice dispersion relation
analysis.
Using these ``surface-corrected
lattice dispersion relation polarizabilities'',
\citet{CoDr04} show that for the case $m=3+4i$,
the error in $Q_{\rm abs}$ can be reduced to of order $1\%$ for
$x=0.83$ using as few as $N=5904$ dipoles, whereas for $x=0.83$, the
error is seen from Figure \ref{fig:m=3.00+4.00i} 
to be $\sim20\%$ even with $N=17904$ dipoles.
Such surface corrections will obviously be of value for DDA calculations,
but at the moment have been applied only to sperical, spheroidal,
or ellipsoidal targets \citep{CoDr04}.

\section{Summary}

The principal results of this paper are as follows:

\begin{enumerate}
\item We have derived the exact mode equation (\ref{eq:mode_equation}) 
for propagation of electromagnetic
waves through a rectangular lattice of polarizable points.

\item If the electric polarizability tensor $\alpha$ for the lattice 
points is
specified, then we can solve the mode equation to obtain 
the dispersion relation for waves propagating
on the lattice.
Alternatively, if we wish the lattice to have the same dispersion relation
as a continuum medium of complex refractive
index $m$,
then we can solve the mode equation 
for the appropriate polarizability tensor.

\item In the case of a cubic lattice, our derived ``lattice dispersion
relation'' polarizability tensor (\ref{eq:alpha_cubic}) is diagonal
but anisotropic, and differs from
the isotropic polarizabilities obtained by DG93.

\item The new polarizability prescription has been tested in DDA
scattering calculations for spheres.
The resulting cross sections have accuracies comparable to those
obtained with the DG93 prescription.
\end{enumerate}
\acknowledgements
We thank Robert Lupton for making available the SM software package.
This work was supported in part by NSF grants AST-9988126 and
AST-0216105.

\begin{appendix}
\section{\label{app:R0R1R2R3}
Numerical Results for $R_0(i)$, $R_1$, $R_2(i)$, and $R_3(i,j)$}

The coefficients $R_0(i)$ are obtained numerically
from the sums (\ref{eq:R_0}).
Note that eq. (\ref{eq:R_0}) for $R_0$ involves the difference between two
sums which separately diverge at infinity.
To handle this divergence, one notes that in nature one is always interested
in finite wavelengths and nonzero frequencies; when dealing with finite
wavelengths, the lattice sums converge because of the oscillations introduced
at large distances by retardation effects.
To obtain the proper behavior when going to the zero frequency limit, we
modify the summations to suppress the contribution from large distances.
The manner in which this is achieved is not critical.
We compute
\beq
R_0(i)=\lim_{\alpha\rightarrow 0}
	\left[
	\sum_\bn
	 { n_i^2 \over |\bn|^2 }\exp(-\alpha |\bn|^4) - 
	\sum_\bn
	 { Q_i^2 \over |\bQ|^2 }\exp(-\alpha |\bQ|^4)
	\right]
\label{eq:evalR_0}
\eeq
The exponential cutoff allows one to limit the summations to
$\alpha|\bn|^4 < 20$ and $\alpha|\bQ|^4 < 20$.
The summation (\ref{eq:evalR_0}) is carried out for various small values
of $\alpha$ and it is found, as anticipated, that
for $\alpha < 10^{-8}$ the numerical result for
$R_0$ is insensitive to the choice of $\alpha$.
Each of the sums in (\ref{eq:evalR_0}) is over a lattice; $|\bn|^2$
defines the distance from the origin for the cubic lattice, and
$|\bQ|^2$ is the distance from the origin for the reciprocal lattice.
Each sum diverges at infinity, where we can replace the discrete summation
by an integration; the reason why the divergences cancel is
that the two lattices have the same density of lattice points, so that
at large distances, both approach the same integration.


\begin{table}[h]
\caption{\label{tab:R_0}
	$R_0(i)$ for selected values of $d_1:d_2:d_3$}
\bigskip
\centering
\begin{tabular}{|c|c|c|c|c|c|}
\hline
$d_1$&	$d_2$&	$d_3$&	$R_0(1)$&	$R_0(2)$&	$R_0(3)$\\
\hline
1&	1&	1&	0&		0&		0\\
1&	1&	1.5&	0.20426&   0.20426&  -0.40851\\
1&	1.5&	1.5&	0.52383&  -0.26192&  -0.26192\\
1&	1&	2&	0.38545&   0.38545&  -0.77090\\
1&	1.5&	2&	0.81199&  -0.21028&  -0.60172\\
1&	2&	2&	1.19693&  -0.59846&  -0.59846\\
1&	1&	3&	0.74498&   0.74498&  -1.48995\\
1&	1.5&	3&	1.38481&  -0.14304&  -1.24176\\
1&	2&	3&	1.96224&  -0.69714&  -1.26510\\
1&	3&	3&	3.11030&  -1.55515&  -1.55515\\
\hline
\end{tabular}
\end{table}


The quantities $R_1$, $R_2$, and $R_3$ appearing in the dispersion relation
in the limit of small but nonzero frequency $\nu$ have been obtained
by direct evaluation of $R_3(i,j)$ using eq. (\ref{eq:R_3}) followed by
use of equations (\ref{eq:R_1fromR_3}) and (\ref{eq:R_2fromR_3}) to obtain
$R_1$ and $R_2(i)$.
Six elements of the symmetric matrix $R_3(i,j)$ are given in Table
\ref{tab:R_3}.

Finally, $R_1$ and $R_2(i)$ are given in Table \ref{tab:R_1andR_2}.

\begin{table}[h]
\caption{\label{tab:R_3}
	$R_3(i,j)$ for selected values of $d_1:d_2:d_3$}
\bigskip
\centering
\begin{tabular}{|c|c|c|c|c|c|c|c|c|}
\hline
$d_1$& $d_2$& $d_3$& $R_3(1,1)$& $R_3(2,2)$& $R_3(3,3)$& $R_3(1,2)$& $R_3(1,3)$& $R_3(2,3)$\\
\hline
  1&  1&  1&  0.00000&  0.00000&  0.00000&  0.00000&  0.00000&  0.00000\\
  1&  1&  1.5&  0.37743&  0.37743& -1.62922&  0.13566&  0.01609&  0.01609\\
  1&  1.5&  1.5&  0.80161& -0.80815& -0.80815&  0.16648&  0.16648& -0.18041\\
  1&  1&  2&  0.55693&  0.55693& -3.72878&  0.23735&  0.09360&  0.09360\\
  1&  1.5&  2&  1.02166& -0.55501& -2.30887&  0.27206&  0.25987& -0.21806\\
  1&  2&  2&  1.26456& -1.78732& -1.78732&  0.37528&  0.37528& -0.39535\\
  1&  1&  3&  0.81662&  0.81662& -8.83412&  0.39793&  0.23223&  0.23223\\
  1&  1.5&  3&  1.34832& -0.40088& -6.06792&  0.43771&  0.42272& -0.15845\\
  1&  2&  3&  1.62638& -1.56624& -4.80661&  0.55590&  0.55480& -0.48211\\
  1&  3&  3&  2.04073& -3.94766& -3.94766&  0.76142&  0.76142& -0.90991\\
\hline
\end{tabular}
\end{table}

\begin{table}[h]
\caption{\label{tab:R_1andR_2}
	$R_1$ and $R_2(i)$ for selected values of $d_1:d_2:d_3$}
\bigskip
\centering
\begin{tabular}{|c|c|c|c|c|c|c|}
\hline
$d_1$&	$d_2$&	$d_3$&	$R_1$&	$R_2(1)$&	$R_2(2)$&	$R_2(3)$\\
\hline
  1&  1&  1&  0.00000&  0.00000&  0.00000&  0.00000\\
  1&  1&  1.5& -0.53869&  0.52918&  0.52918& -1.59705\\
  1&  1.5&  1.5& -0.50962&  1.13457& -0.82209& -0.82209\\
  1&  1&  2& -1.76582&  0.88788&  0.88788& -3.54158\\
  1&  1.5&  2& -1.21448&  1.55359& -0.50100& -2.26706\\
  1&  2&  2& -1.59967&  2.01512& -1.80739& -1.80739\\
  1&  1&  3& -5.47612&  1.44677&  1.44677& -8.36967\\
  1&  1.5&  3& -3.71651&  2.20875& -0.12162& -5.80365\\
  1&  2&  3& -3.48931&  2.73708& -1.49246& -4.73393\\
  1&  3&  3& -4.62875&  3.56356& -4.09616& -4.09616\\
\hline
\end{tabular}
\end{table}
\end{appendix}

\end{document}